# Centrosymmetric PbTe/CdTe quantum dots coherently embedded by epitaxial precipitation

W. Heiss, H. Groiss, E. Kaufmann, G. Hesser, M. Böberl, G. Springholz, F. Schäffler

*Institute for Semiconductor and Solid State Physics, University Linz, Altenbergerstraße 69, 4040 Linz, Austria*

K. Koike, H. Harada, and M. Yano

*Osaka Institute of Technology, Asahi-ku Ohmiya, Osaka 535-8585, Japan*

A concept for the fabrication of highly symmetric quantum dots that are coherently embedded in a single crystalline matrix is demonstrated. In this approach, the formation of the quantum dots is induced by a transformation of an epitaxial 2D quantum well into an array of isolated precipitates with dimensions of about 25 nm. The formation process is driven by the immiscibility of the constituent materials resulting from their different lattice structures. The investigated PbTe/CdTe heterosystem combines two different cubic lattices with almost identical lattice constants. Therefore, the precipitated quantum dots are almost strain free and near thermodynamic equilibrium they exhibit the shape of small-rhombo-cubo-octahedrons. The PbTe/CdTe quantum dots, grown on GaAs substrates, display intense room temperature luminescence at wavelength around 3.2 μm, which makes them auspicious for applications in mid-infrared photonic devices.

Semiconductor quantum dots (QDs), exhibiting size tunable optical and electronic properties which are similar to those of atoms or molecules, are key elements in advanced opto-electronic devices such as high-efficiency lasers,[1,] and single photon sources.[2] The most efficient electronic access to the quantum dots is obtained when they are coherently embedded in a crystalline matrix. Therefore, not only the chosen synthesis method is crucial for the operation of quantum dot devices, but also the quantum dot shape and size transformations during the embedding process. For the most commonly used self-organization process, namely the strained layer heteroepitaxy in the Stranski-Krastanov growth mode[3,4], where coherent nanoislands spontaneously nucleate on the surface of a thin wetting layer, the overgrowth of dot layers basically results in a sequence of shape transitions which is nearly the reverse of those during the growth[4]. Therefore, buried InAs[5] or SiGe[6] dots have the shape of truncated pyramids with large differences in the vertical and lateral extend. Furthermore, strong intermixing with the matrix material is observed when the dots are grown at high temperatures[7] or in ternary quantum dots[8], resulting in a decrease of electronic confinement energies. Spontaneous formation of islands with well defined interfaces are obtained, e. g., by epitaxial growth of alloys with compositions well above the solubility limit for one of the constituent. In this case, precipitation results in coherently embedded islands, which are randomly distributed within the matrix and show irregular island shapes[9]. Highly symmetric quantum dots, in contrast, are formed in glasses[10] and by chemical synthesis of colloidal nanocrystals in liquids[11]. For the first method the matrix is insulating and amorphous and for the latter, coherent embedding in crystals[12] is rather demanding, due to the organic shell of the nanocrystals and the lack of control regarding their crystalline orientation in respect to that of the matrix.

Here we present a method for the fabrication of almost intermixing-free QDs with highly symmetric shapes and height-to-widths ratios close to one, which are coherently embedded in a crystalline matrix. It is based on the heteroepitaxy of a combination of two almost immiscible materials, in our case PbTe and CdTe, followed by a thermal induced precipitation step. Unlike the case of $In_{1-x}Ga_xN$, where the miscibility gap is caused by the difference in lattice constants[13], here it results form the difference in lattice structures[14], since PbTe crystallizes in the rock-salt structure, whereas CdTe exhibits the zinc-blend structure. The dimensions of these two cubic lattices are almost identical, ($a_{PbTe}$=0.6462 nm, $a_{CdTe}$=0.6480 nm)[15] so that the quantum dots are essentially strain free after growth, which is a precondition for the formation of dots with highly symmetric shapes. The large difference in band gap energies of 1.2 eV results in an efficient carrier confinement, making these dots highly luminescent at room temperature. The emission is found at wavelength around 3 μm, i.e., in the mid-infrared spectral region, being important due to many characteristic molecular vibrational-rotational absorption lines.

The quantum dots are formed by two steps: First, single PbTe/CdTe quantum wells are grown by solid source molecular beam epitaxy on standard semi-insulating (001) GaAs substrates. To reduce dislocation densities, 900 nm thick CdTe buffer layers with burrried CdTe/MnTe superlattices were used. Two dimensional growth of the PbTe quantum wells was obtained at rather low substrate temperatures between 220 and 250 °C. The two dimensional growth is evidenced by observing streaks in the in-situ reflection high energy electron diffraction (RHEED) patterns, which are typical for a two-dimensional layer-by-layer growth. The growth of the buffer layers and the quantum well samples with a CdTe capping layer of 50 nm is analogous to that in Refs 15. After gowth, the samples are annealed for 10 min in an inert atmsophere at varios temperatures to transform the quantum well into an array of quantum dots. Thermal post-growth annealing has already been used to form quantum dots from other materials, such as CdSe/ZnSe[16] and $Cd_{1-x}Mn_xTe$[17], but for the PbTe/CdTe material system presented here the results are strikingly different, because of the intrinsic immiscibility of PbTe and CdTe and their almost identical lattice constants.

The samples were investigated by cross sectional transmission electron microscopy (TEM) with a JEOL 2011 FasTEM instrument operated at 200 keV. Figure 1 (a) shows a dark field image of the as-grown sample cross section using the (002) reflection, which is only symmetry allowed in the rock salt lattice. Thus, only the PbTe regions become visible, indicating that the PbTe layer retains its rock salt crystal structure within the zincblende CdTe matrix. For the as grown sample the PbTe quantum well is observed in the form of elongated islands with an average thickness of about 10 nm and lateral extents of several 100 nm. This indicates that the transformation of the initially two dimensional PbTe layer into quantum dots already started, either during the overgrowth with the CdTe cap or during the preparation of the TEM sample, due to an considerable thermal load imposed by the ion milling. The high resolution TEM image of the quantum well in Fig. 1(b) resolves the different atomic configurations of the two lattices. This image was recorded along the [110] zone axis, with the abscissa along [1$\bar{1}$0] and the ordinate along [001], as indicated in the inset of Fig. 1(d). Because of the different structure factors of rock salt and zincblende lattices, the (horizontal) (2$\bar{2}$0) and the (vertical) (002) lattice planes are resolved in the PbTe quantum well, whereas only the {111} lattice planes are symmetry-allowed in the CdTe barrier layers. The image evidences the coherent growth of the two different lattices on each other, and it shows the nearly atomically sharp upper and lower (001) interfaces.

Annealing at $T_a$=350°C results in a complete breaking-up of the PbTe film into well-separated, highly symmetric QDs. Figure 1(c) shows two such dots, which appear in the projection as dark squares with truncated corners. Both, the heights and widths of the dots are in the order of 25 nm. In the high resolution image in Fig. 1(d), the three interface classes

{100}, {110} and {111} around the periphery of the dot are clearly visible. These dot facets suggest that the dots have the highly symmetric shape of a small-rhombo-cubo-octahedron, as sketched in the inset of Fig. 1(c). It is worth to note, that the same dot shapes have previously been demonstrated for colloidal PbSe nanocrystals[18] as well as for PbSe/SiO$_2$ quantum dots fabricated by laser ablation[10]. This indicates, that the scmall-rhombo-cubo-octahedron represents the dot shape in thermal equilibrium, which is formed by minimizing the PbTe/CdTe interface energies. This special shape is obtained, when all three interface energies are equal.

A systematic analysis of cross sectional TEM images of more than 120 dots reveals roughly two different dot classes. (1) Highly symmetric dots with width-to-height ratios close to one, and (2) elongated dots with limited heights of up to 27 nm. The two classes are indicated by the shaded areas in Fig. 2, where the gray cone represents aspect ratios between 0.8 and 1.2. From the statistical analysis the mean dot dimensions are 25 nm in height and 29 nm in width. Since the dots are obtained from a transformation of a 2D layer, the observed elongation in lateral direction together with the limited dot heights indicate that the elongated dots have not reached thermodynamic equilibrium. Therefore, further annealing is expected to result in a homogenization of the size distribution. Irrespective of their aspect ratios, all dots have some properties in common: Their heights exceed the initial quantum well width, all of them show exclusively {100}, {110} and {111} facets, and they all exhibit centrosymmetric shapes. This is exemplarily shown in the three insets of Fig. 2, exhibiting bright field TEM images of quantum dots with various aspect ratios.

A particularly striking feature of the PbTe quantum dots is their very intense photo luminescence emission at room temperature. For these investigations, the samples were excited with a continuous wave 1480 nm diode-laser emitting with a maximum power of 245 mW, and the room temperature emission spectrum was recorded by a liquid-nitrogen-cooled InSb infrared detector mounted to a grating spectrometer. The complete setup was kept under an $N_2$ atmosphere to minimize mid-infrared atmospheric absorptions. Figure 3 (a) shows the measured photoluminescence spectra of the PbTe/CdTe samples at different stages of the quantum-well/quantum-dot transformation. For the as-grown sample (250°C spectrum in Fig. 3(a)), a single PL maximum is observed at a wavelength of 2.9 μm (428 meV). It is blue-shifted by 84 meV (0.7 μm) relative to the emission maximum of a 3 μm thick PbTe reference epilayer represented as dashed line in Fig 3(a). This blue shift is caused by the carrier confinement in the PbTe/CdTe quantum well. Annealing the samples results in significant changes of the PL spectra. First of all, the overall output intensity strongly increases with increasing annealing temperature, as indicated by the arrows in Fig. 3. In particular, a factor of ≈10 enhancement is achieved upon annealing at 320°C. Secondly, already after annealing at 280°C, a well-resolved second PL peak appears at a longer wavelength of λ=3.3 μm. At even higher anneal temperatures this peak further increases and exceeds the 2.9 μm PL intensity of the quantum well (see Fig. 3(a)). This peak arises from the PbTe quantum dots formed by the redistribution of the PbTe quantum wells. Both, the transition energies of the quantum well and the quantum dots are in agreement with model calculations assuming the effective electron and hole masses of CdTe ($m_e^* = 0.094\ m_0$, $m_h^* = 0.42\ m_0$) and PbTe ($m_e^* = 0.043\ m_0$, $m_h^* = 0.051\ m_0$ )[15], as well as a ratio between the conduction and the valence band offsets of 11 to 1. Moreover, the absence of any blue shift of the QW emission at 2.9 μm indicates that no significant intermixing between the PbTe quantum well and the CdTe matrix occurs during the annealing process. As shown in Fig. 3(b), when the PL is excited above the CdTe barriers using a 730 nm Ti:sapphire laser, the emission intensity from the samples is even by a factor of 30 higher than that of the high-mobility PbTe bulk reference sample or any other PbTe based multi-quantum well or superlattice sample tested in our laboratory[19]. This high luminescence yield arises from the peak-like density of states typical for zero-dimensional systems as well as from the efficient carrier trapping in quantum dots due to the large energy barriers. The latter drastically improves the spatial overlap between electron and hole wave functions, whereas for bulk PbTe the extremely high dielectric constant of $\varepsilon_0=400$ suppresses the formation of excitons, in contrast to the situation for wide band gap semiconductors.

Eptiaxial quantum dots with atomically sharp interfaces and centrosymmetric shapes are fabricated by post growth thermal annealing of a single PbTe/CdTe quantum well. Thermal activation leads to a demixing of the two materials, resulting in the precipitation of PbTe quantum dots, which are coherently embedded in the CdTe matrix. The dots are almost strain free due to the similar lattice constants of CdTe and PbTe, and their shapes are close to small-rhombo-cubo-octahedrons with an average size of about 25 nm, with some of the dots being elongated in lateral direction. The dot formation is directly observed in the photoluminescence spectra, which shows distinct peaks for the well and the dots, due to the different carrier confinement energies. The mid-infrared emission from a single quantum dot layer exceeds by far that of a bulk-like PbTe reference, proving the high quality of the PbTe/CdTe heterostructures and making them a potentially interesting material for the development of infrared-optical devices.

This work was supported by the Austrian Science Fund, the START project and IRON special research program and by the GME Austria. Two of the authors (K. Koike and M. Yano) were supported by MEXT Japan (15656100) for a part of this work.

**Figure Captions:**

**Figure 1:** TEM images of PbTe/CdTe heterostructures before and after annealing. (a): Dark field images of the quantum well where the PbTe film corresponds to the bright areas of the image. (b): High resolution bright field image of the same quantum well, where the (002) and (220) lattice planes are resolved in the PbTe and the {111} planes in the CdTe matrix. (c,d): Bright field images of the same sample after annealing at 350°C showing the formed PbTe QDs with the shape of small-rhombo-cubo-octahedrons, as sketchted in (c).

**Figure 2:** Size distribution of the PbTe/CdTe dots determined form cross sectional TEM images. The solid line represents an aspect ration of 1. The two gray areas indicate highly symmetric dots with aspect ratios between 0.8 and 1.2 and elongated dots with heights smaller than 27 nm. The experimental points for the dots shown in the insets are indicated by larger symbols (circles).

**Figure 3:** CW-room temperature photoluminescence spectra of a PbTe/CdTe heterostructures measured after annealing at various temperatures (solid lines) and of a bulk-like PbTe reference sample (dashed lines). In (a), the luminescence is excited energetically below the CdTe barriers with a 1480 nm pump laser at a power of 245 mW and in (b) above the barriers with a 730 nm Ti:sapphire laser with 400 mW. The inset shows the nominal sample structure and the arrows indicate luminescence maxima from the quantum well (QW) and from the QDs (QD).

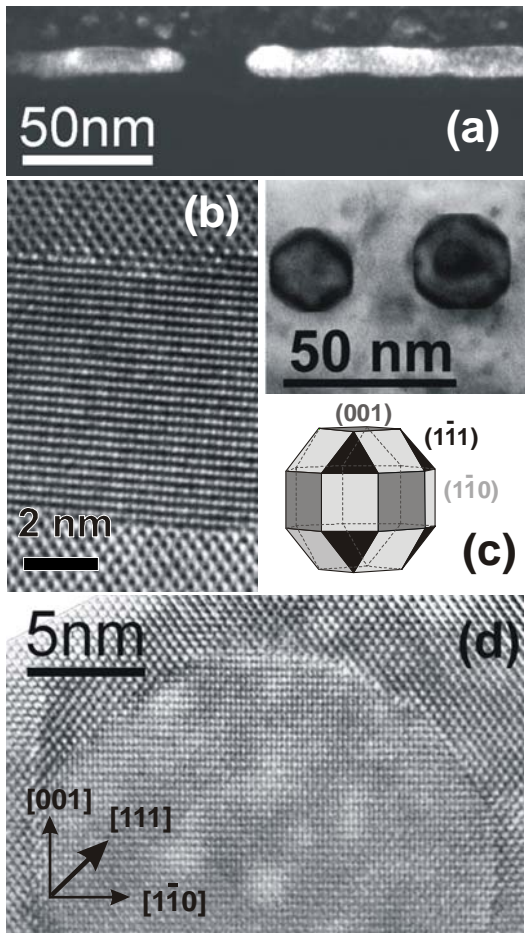

Figure 1 Heiss et al.

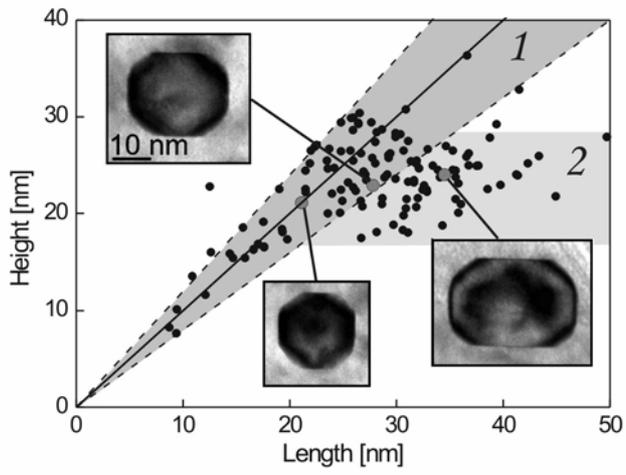

Figure 2, Heiss et al.

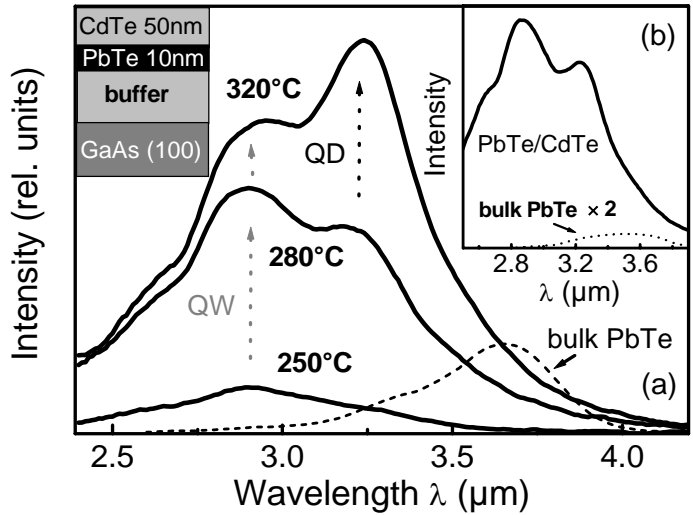

Figure 3: Heiss et al.